# Self assembly of highly fluorescent semiconductor nanorods into large scale smectic liquid crystal structures by coffee stain evaporation dynamics

*Concetta Nobile, Luigi Carbone[§], Angela Fiore, Roberto Cingolani, Liberato Manna, and Roman Krahne\**

National Nanotechnology Laboratory of CNR-INFM, Via Arnesano, 73100 Lecce, Italy.

roman.krahne@unile.it

§Present address: Johannes Gutenberg Universität Mainz, Institute for Physical Chemistry, Jakob Welder Weg 11, D-55128 Mainz

We deposit droplets of nanorods dispersed in solvents on substrate surfaces and let the solvent evaporate. We find that strong contact line pinning leads to dense nanorod deposition inside coffee stain fringes, where we observe large-scale lateral ordering of the nanorods with the long axis of the rods oriented parallel to the contact line. We observe birefringence of these coffee stain fringes by polarized microscopy and we find the direction of the extraordinary refractive index parallel to the long axis of the nanorods.

.

self assembly, semiconductor nanorods, coffee stain, microfluidics

Colloidal nanocrystals represent an interesting material system with great potential in many applicative fields. One crucial prerequisite for the realization of useful structures at macroscopic level is the self assembly of nanocrystals into ordered arrays (for an overview see [1, 2]). Nanorods are particularly challenging in this respect since their shape anisotropy leads to different types of oriented assemblies resulting in diverse physical ensemble-properties. Recently, significant progress has been made on self assembly methods that are based on nanorod deposition from solution, and on the understanding and mastering of approaches that exploit different forms of interactions.[3-5] The application of external electric fields, for example, leads to great control over the alignment direction and to micron size areas of oriented nanorod arrays.[6-11] Other approaches involve fluid flow dynamics of the solvent, such as slow solvent evaporation on a liquid-solid-air interface,[12] solvent fluidics and the presence of liquid-air interface in the formation of a lyotropic phase from a drying solution,[13] and inter-particle interactions as well as rod solubility in a binary solvent/non-solvent liquid mixture.[14-21]

We report here the self assembly of colloidal CdSe/CdS core-shell nanorods into laterally oriented arrays by means of the fluid dynamics of drying droplets. In a drying droplet the pinning of the contact line leads to a radial flow of fluid from the center to the edge of the drop in order to compensate for the evaporated liquid at the surface. If the liquid contains nanoparticles this flow leads to a significantly enhanced density of nanoparticles near the contact line, and when the contact line eventually retreats the so called "coffee stains" are formed, i.e. fringes of densely packed nanocrystals.[22-24] If a droplet of a solution of toluene containing CdSe/CdS nanorods evaporates, we observe the formation of large ribbon structures in which the nanorods are aligned side by side within the fringes of the coffee stains, whereas in the central regions we find no specific ordering of the nanorods. For solutions with high rod density the pinning of the contact line is enhanced and we obtain large scale alignment of nanorods in smectic liquid crystal structures within the dense fringes, where the long axis of the nanorods is oriented along the contact line. We find that also external electric fields and surface patterning can influence the pinning of the contact line and its retraction motion, which can be exploited in order to achieve ordered assemblies of nanorods. We use polarized microscopy to characterize the refractive properties of the obtained nanorod assemblies.

In the coffee stain effect fringes consisting of deposited material form due to pinning and subsequent retraction of the contact line of a drying droplet. The pinning of the contact line induces an outward flow of the liquid from the inner regions towards the edge of the droplet in order to compensate for the evaporated solution from the droplets surface. The radially outward flow transports particles that are dissolved in the solution towards the contact line of the droplet. When the contact line eventually retracts, it leaves a ring-like structure of dense material on the surface which is called the coffee stain.

In this work we investigate the coffee stain effect and the resulting patterns for solutions of nanorods dispersed in toluene. We find that the assembly pattern of the nanorods depends significantly on the nanorod density in the solution and on the strength of the contact line pinning. We start with discussing our experiments for solutions with nanorod concentrations below $10^{-6}$ M. We deposited small droplets of toluene (10 – 30 µl) with different nanorod concentrations ( 6 x $10^{-7}$ M, 6 x $10^{-8}$ M, 1 x $10^{-8}$ M) on a $SiO_2$ substrate and let the solvent evaporate under ambient conditions. We find that the contact line retracts, undergoing pinning and depinning, in an overall concentric motion, although locally some "arch-shaped" irregularities and fingering occur as they were observed in Ref.[22]. After drying we observed fringes on the substrate consisting of densely deposited nanorods, as shown in Figure 1b. We find ribbon-like nanorod assemblies within the dense fringes in which a bunch of nanorods are aligned side-by-side, as displayed in Figure 1d, whereas in the area in between the fringes the rods are deposited in a layer of random order (Figure 1c). The nanorod ribbons displayed in Figure 1d form "spaghetti-like" entangled structures, and we find that with decreasing nanorod concentration in the solution the length and the density of the ribbons in the fringes decreases (see Figure S2 of Supp. Inf.) Similar ribbon structures formed by nanorods have also been observed for hard, rodlike ß-FeOOH suspensions [25-27], Se and CdSe nanorods [17, 28], and also for gold nanorods[29].

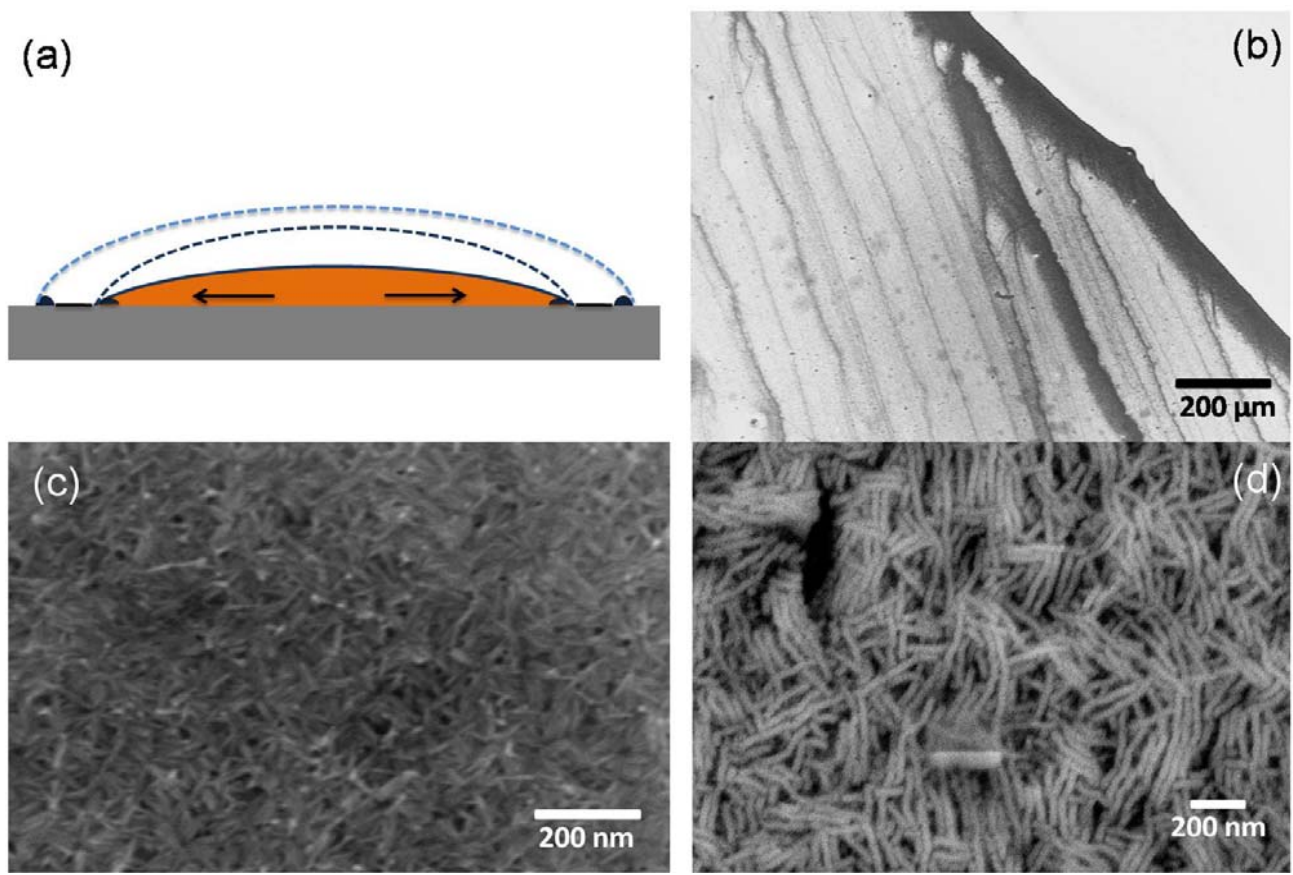

*Figure 1: (a) Sketch illustrating the evaporation and convection processes in a fluid droplet. The arrows indicate the outward flow that compensates for the evaporated liquid while the contact line is pinned. The dashed lines illustrate the drop-air interfaces previous to the solvent evaporation. (b) Low magnification SEM image showing part of a coffee stain with fringes consisting of nanorods, (c) larger magnification SEM image of the region in between the fringes where the rods show no specific ordering. (d) SEM images of an area within a fringe where the nanorods assemble laterally side-by-side into ribbons. For transmission electron microscopy images of the nanorod sample see Supp. Inf. Figure S1 a.*

Figure 2 shows what happens if we deposit droplets of toluene solutions with nanorod concentrations of $10^{-6}$ Mol on $SiO_2$ or glass. Here SEM inspection of the coffee stain fringes (Fig 2a) shows that the nanorods assemble in much higher order. Higher magnification images show (see Fig 4) that the nanorods are aligned in smectic structures with their long axis parallel to the circular fringe. Within the fringe we find stripes with differing density. We observe that the inner border of the stripe with the highest density (see dotted lines in Fig 2a and b) is well defined and that this part of the coffee stain shows the highest order of the nanorod assembly. We find that the increased nanorod density in the solution leads to stronger pinning of the contact line and a thicker fringe of deposited nanorod material within the fringe. The thick fringe of nanorod material prolongs the pinning of the contact line and, subsequently, leads to the rupture of the liquid film behind the fringe (see Fig. 4c), similar to the mechanism reported in Ref. [30]. Following this model, the ring of deposited material behind the contact line prolongs the pinning until the minimum of the droplet height behind the deposit leads to a rupture of the fluid film.

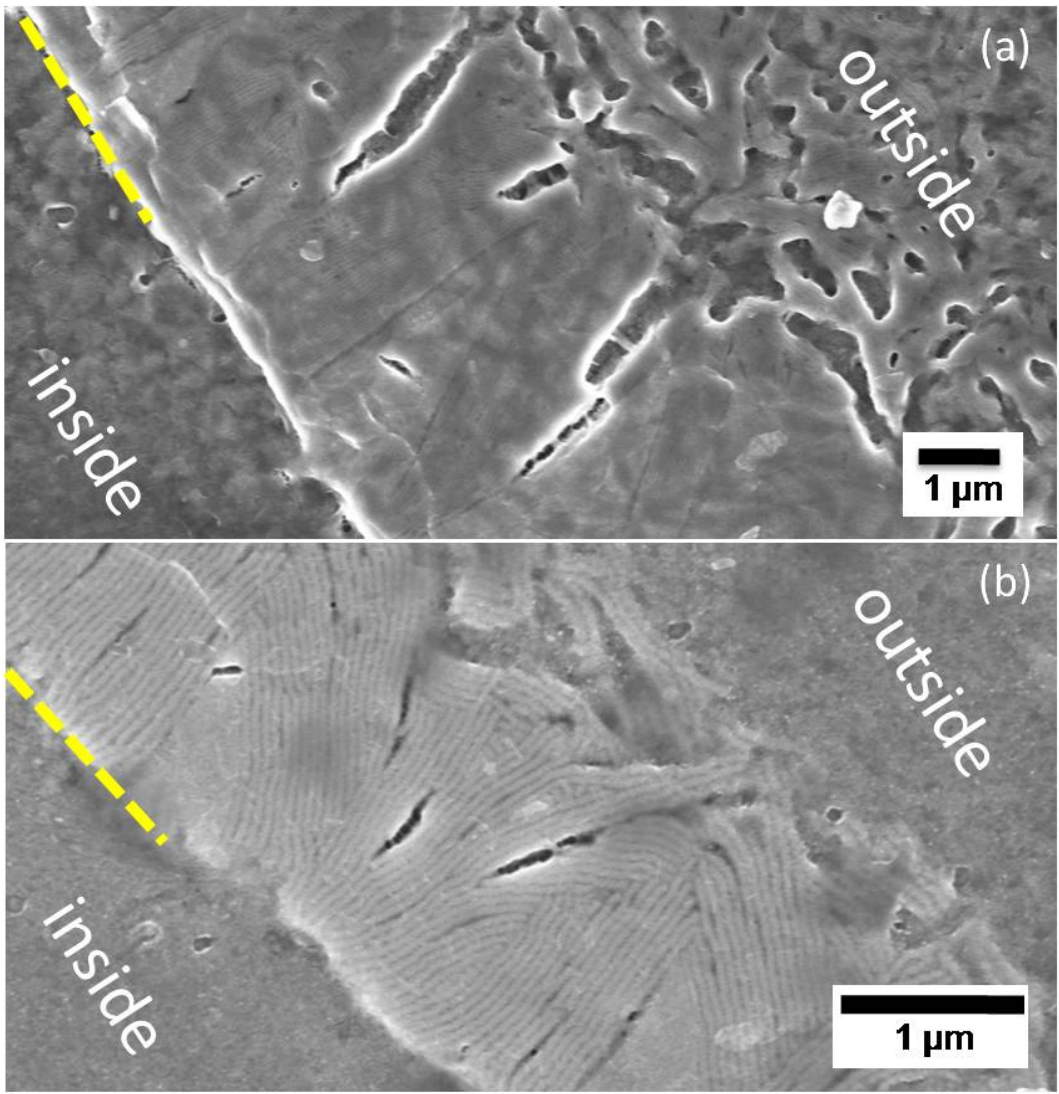

*Figure 2: SEM images of the coffee stain fringe for nanorod deposits with higher (a) and lower (b) density. We observe high lateral order with the individual nanorods aligned parallel to the contact line. We note the well defined inner border of the dense stripe of the fringe and the large degree of lateral rod alignment near this border.*

The nanorods in the coffee stain fringes shown in Fig. 2 align parallel to the contact line. This arrangement results from the balance of the torque exerted on the ends of the nanorods by the solvent flow towards the pinned contact line. In this picture the first rods that encounter the pinned contact line orient along the contact line and act as nucleation centers for the following parallel arrangement. We know from previous studies [10] that the CdSe/CdS core/shell rods arrange easily into ordered lateral arrays. Figure 3 shows SEM images where the nanorods align parallel to the border line of voids. Here the nanorods were dissolved in toluene and GaAs substrates were immersed at a tilted angle of 45° in the nanorod solution and the solvent was allowed to evaporate. We find that the tendency of the rods to align parallel to the border line and the smectic order of the deposited nanorods increase with increasing rod density. Theoretical modeling of the evaporation dynamics of solvents containing colloidal nanocrystals show that the voids start to form well before the complete evaporation of the solvent [31]. Consequently, also in this case the torque argument for the nanorod that encounters a contact line applies. We note that the case of the pinned contact line in our coffee stain experiments differs significantly from a continuously retracting contact line where the rods align along the flow direction due to capillary forces.

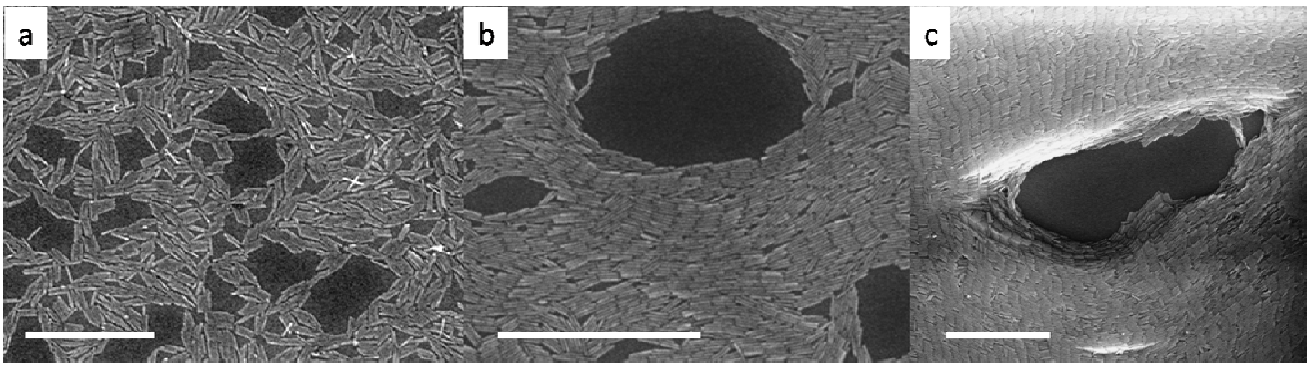

Figure 3: SEM images of CdSe/CdS core/shell nanorods deposited on tilted GaAs substrates by slow solvent evaporation (of the order of 5 nm/s) for increasing nanorod density from a-c. The scale bar corresponds to 500 nm for all images.

We find that the nanorod alignment in the coffee stain region depends mainly on two factors. The first is the nanorod density in the solution. Higher nanorod concentrations lead to better lateral nanorod alignment, i.e. higher degree of order and larger areas of aligned rods. The second factor is the retreating motion of the contact line. Overall concentric retraction of the contact line shows effects like fingering which reduce the degree of order of the nanorods since leads to ruptures of the dense nanorod deposits. However, a circular retractive motion of the fluid, as it was reported in ref. [30] has much less destructive impact on the coffee stain ring and thereby on the order of the nanorod assembly.

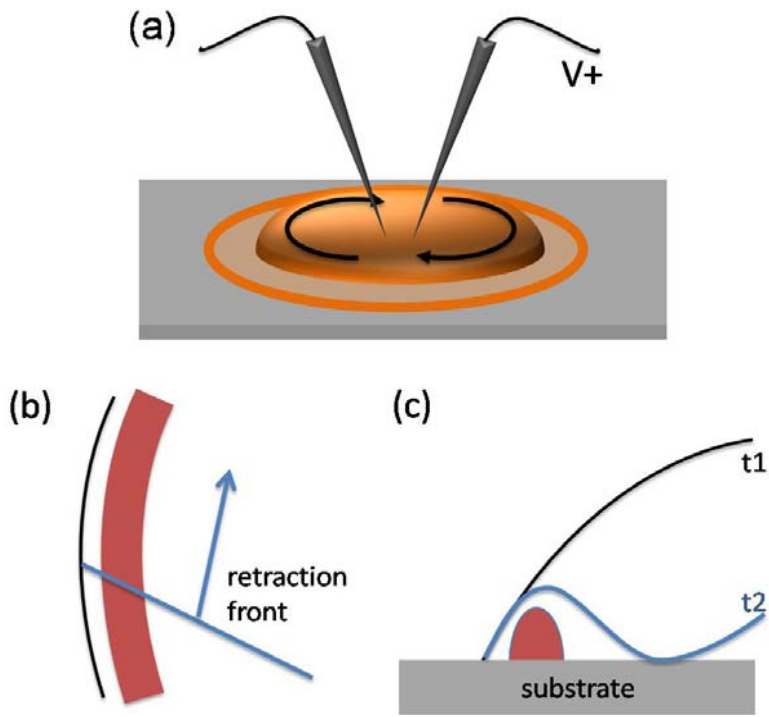

*Figure 4: (a) Applying an external dipole field enhances the pinning of the contact line and the causes circular retraction motion of the fluid. (b) Schematic illustration of the circular motion of the retracting fluid. The red stripe symbolizes the region of densely deposited nanorods. (c) Sketch of a side view of the drop edge. Nanorods are deposited closely to the contact line due to the radially outward fluid flow.*

*The nanorod deposit enhances the pinning of the contact line [30]. We attribute the circular retraction motion of the contact line to a minimum in drop height behind the dense fringe of nanorods (illustrated by the blue line).*

We can stimulate such a circular fluid retraction by applying an electrical dipole field at the center of the drop, as illustrated in Fig 4a. If we position two electrical micromanipulators on the surface at the center of the drop we find that the contact line pinning related to the outermost coffee stain is enhanced. After the rupture of the fluid film we observe a circular retraction of the fluid, as illustrated in Fig. 4b (see Fig. S3 of Supporting Information). Two films showing the drop drying dynamics with, and without, voltage bias of 150 V is available online at (URL needs to be inserted).

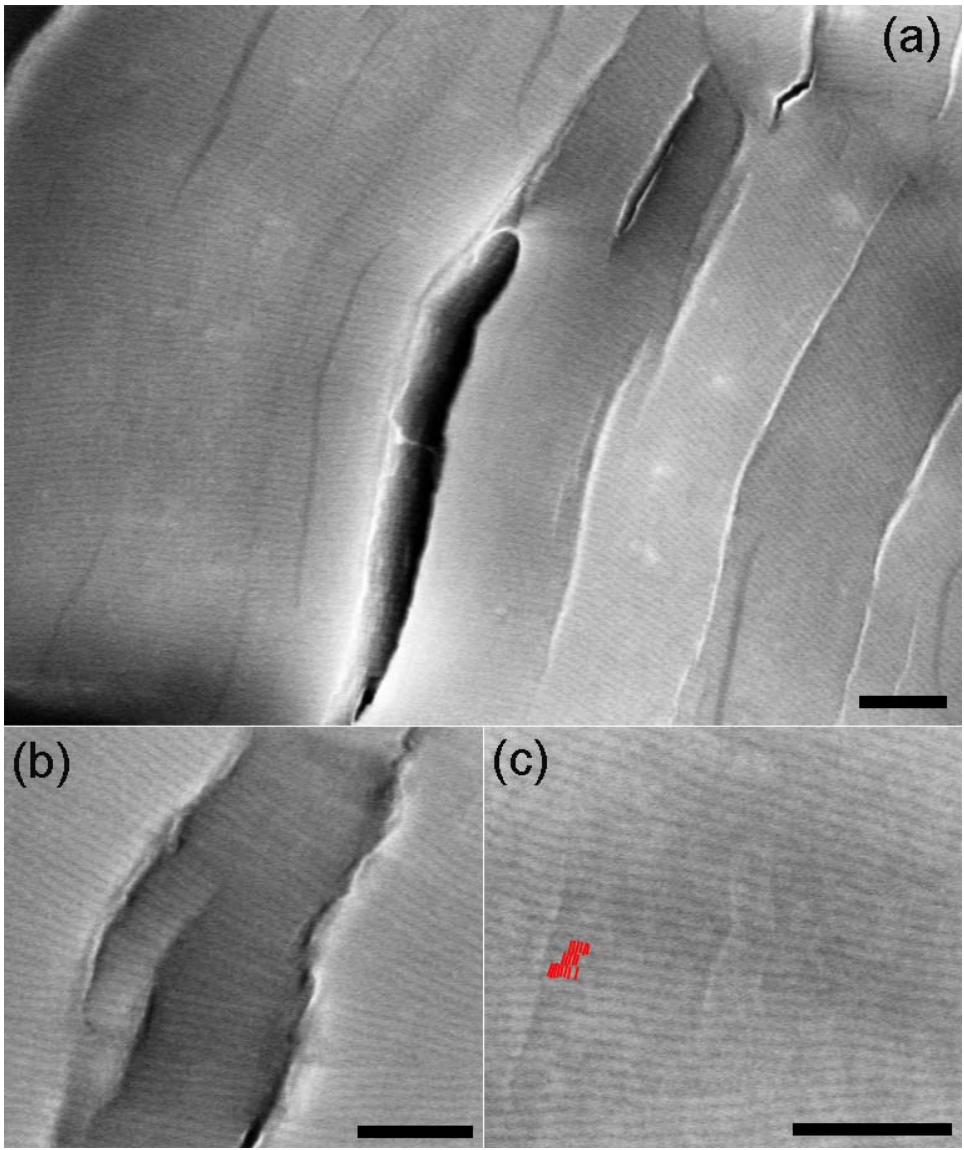

*Figure 5:* *(a-c) SEM images of areas within the coffee stain fringes showing large scale assembly of nanorods in liquid crystal phase forming multilayered stripe patterns. (b) Higher SEM magnification of stripes with different thickness, i.e. different number of layers. (c) Magnification which resolves individual rods within the nanorod ribbons. The scale bar for all three images corresponds to 500 nm. Some nanorods are highlighted in red as a guide to the eye.*

Figure 5 shows the nanorod alignment that we obtained by positioning two electrical micromanipulators (tungsten needles) with 150 V voltage bias at the center of the drop as illustrated in Fig 4a. We find multilayers of nanorods that are aligned in a smectic phase, but in addition the smectic domains arrange in stripes with almost constant width over several micrometers in length. The width of these stripes ranges from some hundreds of nanometers to some microns. Figure 5 b-c shows larger magnifications SEM images that resolve the ribbons and the individual rods within the ribbons. We note that these stripe-like superstructures have relatively sharp, continuous boundaries (less than 50 nm across) that extend over several microns. Patterns like the ones displayed in Figure 5 can extend up to the millimeter scale as shown in Figure 6. We think that these stripe like patterns can be attributed to pinning and de-pinning of the contact line on a comparatively short length scale that resemble a stick-slip movement [30]. We like to mention that the pinning of the contact line can also be influenced by patterning the surface of the substrate. For example, we evaporated a ring shaped Au layer (thickness 30 nm) with a diameter of 5 mm on a glass substrate and found that the contact line pinning was enhanced on the Au a short distance from the Au/glass border. Also in this case we observed the highly ordered lateral assemblies of nanorods.

For samples with a uniaxial anisotropy two refractive indices can be assigned, the ordinary refractive index $n_o$ with orientation perpendicular, and the extraordinary index $n_e$ with orientation parallel to the axis of anisotropy, respectively. The birefringence is then $B= |n_e-n_o|$, and the sign of birefringence can be evaluated using polarized microscopy by inserting a first order retardation plate (530 nm) in the optical path in crossed polarization, with the axis of anisotropy of the sample aligned parallel to the slow axis of the retardation plate. In this configuration positive birefringence leads to blue color, and negative birefringence to red color in the polarized microscopy image.

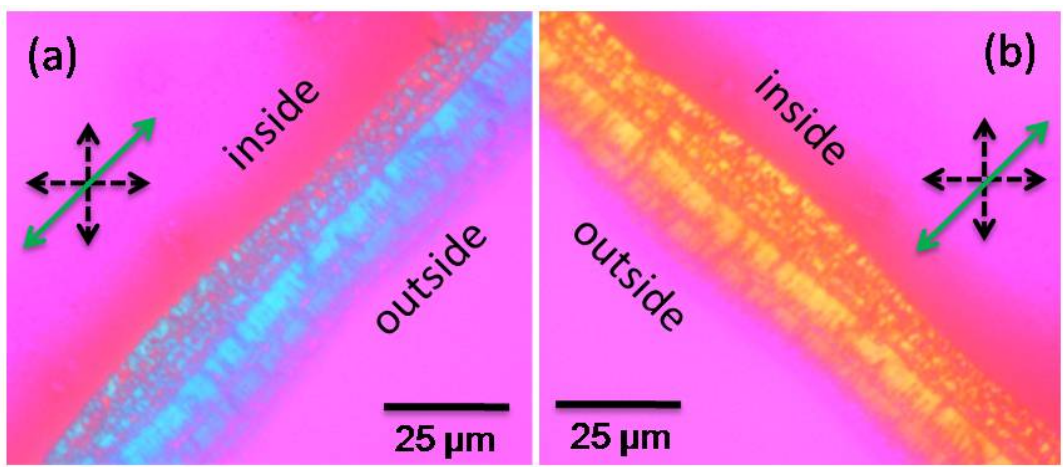

*Figure 6: Polarized microscopy images taken in crossed polarization with a 530 nm retardation plate inserted in the optical path at 45°. The sample was prepared by drop evaporation with 150 V bias voltage as described in the text. The sample has been rotated by 90° from (a) to (b). The dashed black arrows indicate the orientations of the polarizer and analyzer, and the green arrow shows the orientation of the 530 nm retardation plate. "inside" and "outside" refer to which side of the fringe faces the inside/outside of the coffee stain.*

Figure 6 shows part of the outer coffee stain fringe where a bias voltage of 150 V was applied to the manipulator tip during solvent evaporation (as described above). In this case we used a microscope glass slide as substrate. We clearly observe the dense part of the fringe in blue color revealing macroscopic ordering of nanorods arranged in liquid crystal phase. Rotation of the sample by 90° leads to red/orange color of the fringe as it should (Figure 6b). We conclude that the direction of $n_e$ is oriented parallel to the contact line, and from comparison with SEM images we find that $n_e$ is directed along the long axis of the nanorods.

In conclusion, we formed highly ordered multi layers of laterally organized nanorod assemblies inside the coffee stain fringes of an evaporating droplet of dense nanorod solution. In detail, we observed that the individual nanorods form ribbons and that these ribbons arrange in smectic order in the dense central region of the coffee stain fringe. We found that the individual nanorods in this arrangement are oriented parallel to the contact line and we could assign the slow axis of refraction of the nanorod assemblies also along this direction.

Acknowledgements:

The authors thank Benedetta Antonazzo for technical assistance in the nanochemistry lab. The work was supported by the SA-NANO European project (Contract No. STRP013698).

**Supporting information**

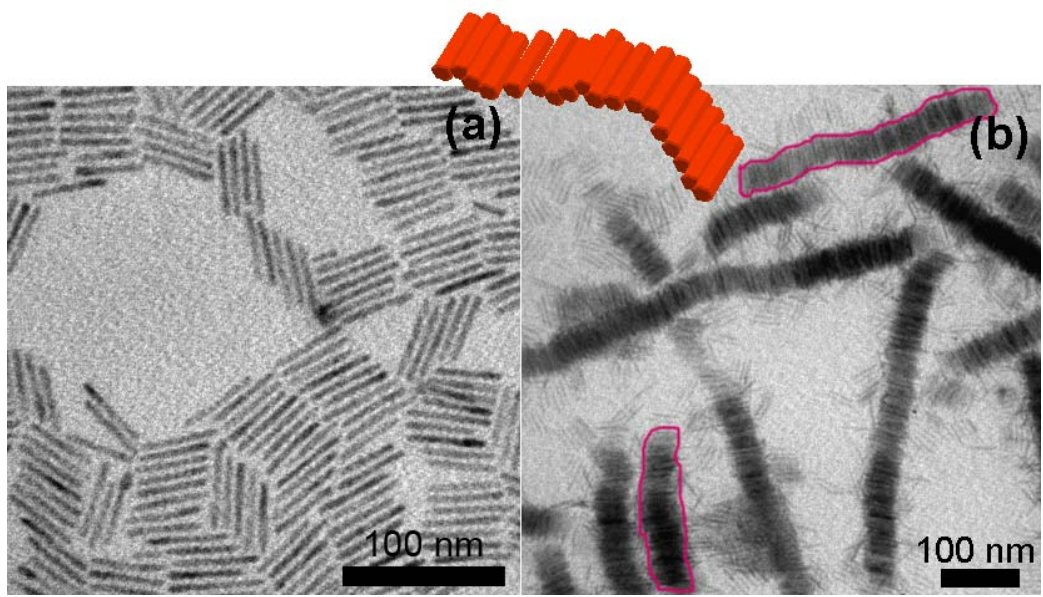

*Figure S1:(a) TEM image of one of the CdSe@CdS nanorod sample (diameter 4 nm, length 50 nm) used for the assembly experiments. (b) TEM images of the nanorod ribbons obtained by drop casting the solution on a silicon-nitride membrane. Some ribbons are highlighted in red as a guide to the eye.*

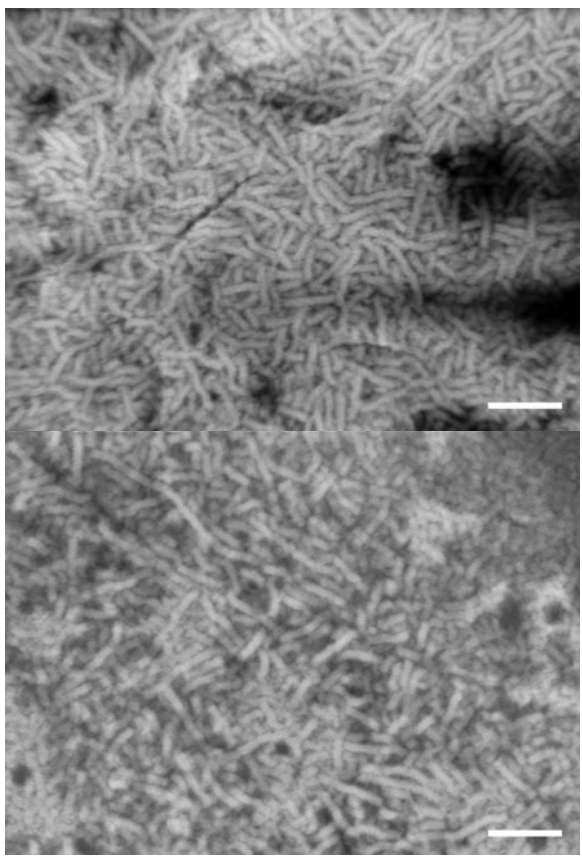

*Figure S2: SEM images of nanorod ribbons observed within the coffee stain fringe. The ribbon length and density depend on the nanorod concentration in the solution (upper figure 6 x $10^{-8}$ M, lower figure 1 x $10^{-8}$ M). The scale bars are 400 nm for both images.*

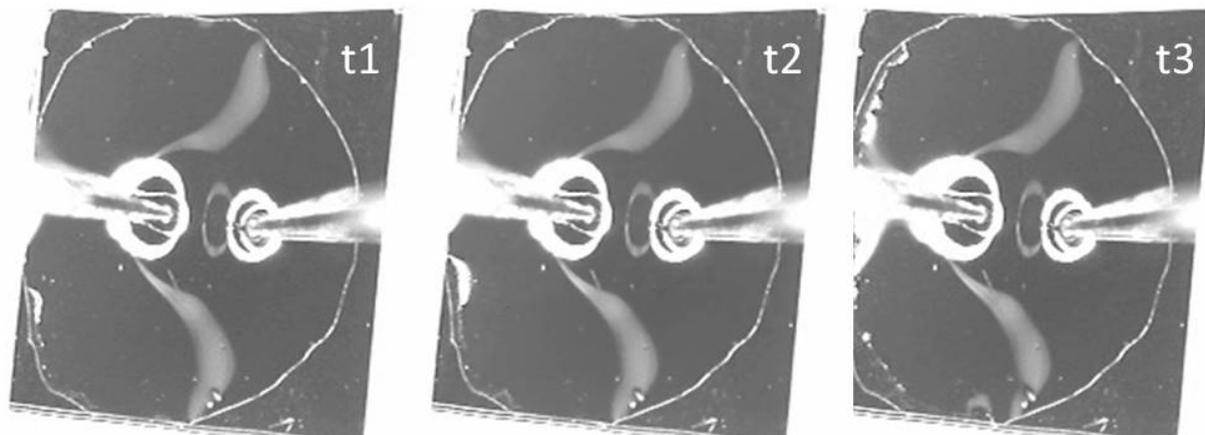

*Figure S3: Microscope images showing a drying drop of nanorod solution at different times. The circular movement of the retreating contact line, proceeding from the lower left section to the upper left section, can be seen. The nanorod concentration was $10^{-6}$ M and the applied bias voltage was 150 V.*